\documentclass[twocolumn,preprintnumbers,amsmath,amssymb,floatfix,showkeys]{revtex4}

\usepackage{graphicx}
\usepackage{dcolumn}
\usepackage{bm}
\usepackage[german,american]{babel}

\newcommand{\mcpcite}[1]{\raisebox{1ex}{\scriptsize \cite{#1}}}

\begin{document}

\title{Understanding Segmental Dynamics in Polymer Electrolytes: A Computer Study    
 }   

\author{Arijit Maitra}
\author{Andreas Heuer}
\email{arijitmaitra@uni-muenster.de, andheuer@uni-muenster.de}

\affiliation{
Westf\"{a}lische Wilhelms-Universit\"{a}t M\"{u}nster, Institut f\"{u}r Physikalische Chemie,
Corrensstr. 30, 48149 M\"{u}nster, Germany }
\affiliation{
NRW Graduate School of Chemistry, Corrensstr. 36, 48149 M\"{u}nster, Germany }

\date{\today}


\begin{abstract}
We study the segmental dynamics of poly(ethylene oxide) (PEO) from microscopic simulations in the neat polymer and a polymer electrolyte (PEO/LiBF$_4$) by analyzing the normal modes. We verify the applicability of the Rouse theory, specifically for the polymer electrolyte where dynamic heterogeneities, arising from cation-polymer interactions, alter the mobility non-uniformly along the chains. We find that the Rouse modes for both the systems are orthogonal despite the presence of non-exponential relaxation of the modes and violation of the Gaussian self-similarity of the chains. The slowdown of the segmental dynamics in the polymer electrolyte is rationalised by an order of magnitude increase in the friction coefficient for those monomers which are bound by cations. In general, for the electrolyte the Rouse predictions for the dynamics of segments (both free and/or bound) agree well except for very short times.

\end{abstract}

\maketitle

\section* {Introduction}

Polymer electrolytes\mcpcite{GraySPE,Ratner} are obtained by dissolution of a salt in a polymer matrix. The multitude of interactions occurring in these systems gives rise to complex dynamical processes of the ions and the polymer segments over a broad range of length and time scales. On the one hand, comprehending the relevance of the mechanisms of ion dynamics will allow the prediction of ionic mobilities, and, on the other hand, the underlying polymer dynamics attains significance due to the coupled nature of the motion of ions and polymer segments.

Computer simulations have provided precious insights on the structural aspects as well as the dynamical processes in polymer electrolytes.\mcpcite{Plathe,Neyertz95,BoroLiIStatic,BoroLiIDynamic,BoroMDLiBF4,BoroMDLiTFSI}
The cations of the salt are solvated by atoms or groups located on the polymer host by means of complexation. This results in a local increase of the friction coefficient of the complexed monomers whereas the uncomplexed monomers are relatively free. The difference in the dynamics of the complexed and the uncomplexed monomers lead to dynamic heterogeneities.\mcpcite{RohrSpiess}
Molecular dynamics (MD) simulations\mcpcite{Allen} of lithium salt in poly(ethylene oxide) (PEO) have shown that the dynamics of oxygen atoms which are bound to a lithium cation (Li$^+$) are slower than those which are free.\mcpcite{MHprl}
The mean square displacements (MSD) of oxygen atoms exhibit a time exponent of
$t^\alpha$ with $\alpha \approx 0.6$ for intermediate times less than the relaxation time of the end to end vector of the polymer.

The widely studied Rouse theory,\mcpcite{Rou53,DoiEd}  known to approximately describe the dynamics of linear, unentangled polymers under melt conditions, capture the essence of the dynamical behavior of the segments. The study of polymer dynamics can be broken down into several dynamical modes with
characteristic time and length scales. Within the framework of the Rouse theory, each mode corresponds to an eigenvector (also called normal coordinate). The eigenvectors are orthogonal to each other, i.e., the different modes are statistically uncorrelated. Knowledge of modes can be employed to calculate the segmental dynamics. The Rouse theory predicts for the segmental MSD a
dependence of $t^{0.5}$ for times $t \ll t_R$ where $t_R$ is the Rouse time or the longest relaxation time of the polymer. 

Very recently, it has been shown for PEO/LiBF$_4$ that it is possible to predict the cationic diffusivity as a function of the microscopic time scales characterising the cation transport mechanisms by considering the coupling of the cationic motion to the segmental dynamics of the polymer.\mcpcite{MHprl} An important delineation was the description of the ether oxygen (EO) dynamics according to the Rouse theory, which was verified by comparison against the MSD of EO atoms, and particularly to account for the retardation of those monomers which are bound to the Li$^+$ ions.

In this paper we elucidate the polymer dynamics from microscopic simulations of amorphous PEO and PEO/LiBF$_4$ and check the extent to which the dynamics can indeed be characterized by the Rouse model. Furthermore, we present Brownian dynamics\mcpcite{DoiEd} simulation of a bead-spring model with heterogeneous friction and some stiffness to closely resemble our chemically realistic polymer electrolyte.
In this way we analyze to which degree the dynamic properties of the microscopically complex PEO/LiBF$_4$ system can be understood from an appropriately chosen bead-spring model.

\section* {Models and Methods}

We have performed MD simulations of a system of i) neat PEO and ii) a model polymer electrolyte composed of PEO and a salt, LiBF$_4$. The concentration of the polymer electrolyte is 20:1 in terms of the EO:Li$^+$ ratio. The Gromacs\mcpcite{Gro} simulation package was used for generating the MD trajectories. All the atoms $-$ viz. carbon, oxygen, hydrogen, lithium, boron and fluorine $-$ were explicitly taken into consideration for setting up the respective systems. The nonbonded interactions between all pairs of atoms including those of the cations and anions were modeled according to the Buckingham potential plus the electrostatic interactions. The two-body effective polarizable force fields employed have been adopted from ref.\mcpcite{PEOFF2003,BoroMDLiBF4} which were derived from quantum chemistry based techniques. The simulation methodology and details of the force fields used in this work closely follows that of ref. \mcpcite{PEOFF2003,BoroMDLiBF4} 

Coulomb interactions between the partial charges (ions as well as polymers) were calculated via the particle-mesh Ewald method with a distance cut-off of 10 \AA. Overall charge neutrality of the systems were also maintained.

The simulation box
contained 16 chains of PEO, with the sequence H-(CH$_{2}$-O-CH$_{2}$)$_{N}$-H, each comprising $N=48$ number of monomers. All bond lengths were constrained to their equilibrium values by the LINCS algorithm\mcpcite{Lincs}. However, the bend and the dihedral angles of the polymer chains were kept flexible.

The simulations were propagated at a constant temperature of \mbox{450 K} in the canonical ensemble and under the application of periodic boundary condition. Temperature coupling was accomplished using the Nos\'{e}-Hoover thermostat.\mcpcite{Nose,Hoover} The average pressure during the runs were of the order of 1 MPa. Equilibration times of the order of $\tau_R$ had been considered for each of the systems prior to the analysis of the trajectories.

Apart from the realistic systems described above, we have also simulated a toy model of a single chain consisting of only
beads and springs, characterized by the standard harmonic
potential. We attempt to capture the essential dynamics of the monomers of our realistic polymers through the toy model. To incorporate correlations between the bond vectors we
impose a bending potential such that the total Hamiltonian is written
as
\begin{equation}
 \begin{split}
 U(\{\mathbf{r}\})  &= \frac{k}{2} \sum_{i=0}^{N-2} (\mathbf{r}_{i+1} - \mathbf{r}_i)^2 \\
            & \quad  - \, k_{\theta}\sum_{i=1}^{N-2} \left\{\frac{(\mathbf{r}_{i}-\mathbf{r}_{i-1})(\mathbf{r}_{i+1}-\mathbf{r}_i)}{\lvert
 \mathbf{r}_{i} - \mathbf{r}_{i-1} \rvert \lvert \mathbf{r}_{i+1}-\mathbf{r}_i \rvert } \right\}
 \end{split}
\end{equation}
where $\mathbf{r}_i$ is the position vector of bead $i$; $k$ and $k_\theta$ are the force constant of the harmonic springs and the bending constant between the adjacent bond vectors, respectively. The forces arising from the bending potential essentially model a freely rotating chain (also see \mcpcite{Winkler}). 
A value of $k=3k_BT$ is chosen, where $k_B=1$ is the Boltzmann constant in reduced units and $T=1$ is the temperature, also in reduced units, used for coupling between the viscous drag and the Gaussian white noise, that entails the mean square bond length $l^2=1.0$. We fix $k_\theta = 2k_BT$ to maintain consistency with the stiffness of the PEO chains in the atomistic PEO/LiBF$_4$ system, yielding a similar value of the characteristic ratio $C_{\infty}$ for both the models; see Table 1.

We construct three different toy systems with system size $N=48$: (i)
BD1: the chain is homogeneous, i.e. all the beads have the same friction coefficient of $\zeta_1=1$, and akin to the characteristics of neat PEO. (ii) BDRA: 12 beads are chosen {\it randomly} with the friction coefficient $\zeta_2=8$ to reflect heterogeneity (or defects). (iii) BDCO: two sets of 6
{\it contiguous} defects are chosen such that the distance between any two particles belonging to different sets is at least 6 particles apart. This realization qualitatively resembles contiguous monomers ($\approx$ 5-7 in number\mcpcite{BoroMDLiBF4}) of PEO bound by cations as in the 20:1 PEO/LiBF$_4$ system. The choice of the friction coeficient of the defect $\zeta=8$ has been rationalised later in this article.

The Brownian dynamics method is employed to evolve the toy systems with a time step of 0.001. For data analysis, all the computed quantities from the toy systems are averaged over 20 independent runs. For BDCO the distance between the two
sets of contiguous defects is chosen randomly for each run.

\section* {Chain Static Characteristics}
 
The global chain properties are listed in Table 1. The correlation between adjacent bond vectors measured as $\langle \cos\theta \rangle$ where $\pi-\theta$ is the angle included between the two adjacent bond vectors, $\mathbf{r}_i-\mathbf{r}_{i-1}$ and $\mathbf{r}_{i+1}-\mathbf{r}_i$, is consistent between the atomistic systems and the toy models. For PEO, $\pi - \theta$ is the angle between the link vectors formed by connecting the ether oxygen atoms consecutively.
The stiffness of the PEO chains as measured by $\langle \cos \theta \rangle$ is slightly higher in the neat system when compared to PEO/LiBF$_4$. This implies
the chains are expanded more for the neat PEO than those in the PEO/LiBF$_4$ system. 

Alternatively, this is also apparent from the values of the characteristic ratio, $C_\infty$=$R^2/l^2(N-1)$,\mcpcite{Flory} where $R^2$ is the mean square end to end distance of the polymer chain, $l^2$ is the mean square bond length and $N-1$ is the number of bonds. Note, that in ref. \mcpcite{SmithPEO} $C_{\infty} \approx 5.1$, for the case of neat PEO, because $l^2$ was defined as the mean square length of a skeletal bond. We consider $l^2$ to be the mean square $O-O$ distance between the adjacent monomers. The characteristic ratio is not sufficient to describe the chain stiffness because it takes into account only the large scale chain conformations. Even though $C_\infty$ is higher for neat PEO when compared to the chains in PEO/LiBF$_4$, the chains in the former display faster relaxation ($\tau_R \approx$ 10 ns) than those in the electrolyte ($\tau_R \approx$ 19 ns). This underscores the importance of local stiffness effects arising from dynamic heterogeneities which are difficult to incorporate in analytical models. Structurally, as evident from the $O-O$ radial distribution function (not shown) and also from the mean square distance between adjacent oxygen atoms, $l^2$, there will be slight differences between the PEO chains in the two systems in terms of the distribution of the conformational states of the $-C-O-$ and $-C-C-$ dihedrals.\mcpcite{SmithPEO,BoroLiIStatic}  
For the toy models all the static properties, which are independent of friction coefficients of the beads, are similar as is expected.

\begin{table}
  \begin{center}
     \begin{tabular}[c]{|c|c|c|c|} \hline
       System       & $l^2$                &  $C_\infty$   & $\langle \cos\theta \rangle$  \tabularnewline[1mm] \hline
       PEO          & 10.0 [\r{A}$^2$]     &  3.3  & $-$0.63       \tabularnewline[1mm] \hline
       PEO/LiBF$_4$ & 9.2  [\r{A}$^2$]     &  2.7  & $-$0.55       \tabularnewline[1mm] \hline
       BD1,BDRA,BDCO          & 1.0                  &  2.6  & $-$0.53       \tabularnewline[1mm] \hline
     \end{tabular}
   \caption{\label{tabsummary} Chain static properties of the different model systems from simulations. The units for the microscopic simulations (PEO and PEO/LiBF$_4$) are provided in box brackets and data for the toy models (BD1, BDRA, BDCO) are expressed in reduced units. ($l^2$: mean square bond length between monomers/segments; $R^2$: mean square end-end vector of the chains; characteristic ratio $C_\infty = R^2/l^2(N-1)$; $\langle \cos\theta \rangle$: mean of the cosine of bond angles).
   }
  \end{center}
\end{table}

\section* {Analysis of Rouse Modes}

\subsubsection* {Statics}
The static and dynamic properties of the polymer chains can be studied by means of the normal mode analysis. The normal modes (also called Rouse modes) for a discrete polymer chain\mcpcite{Verdier} is expressed as
\begin{equation}
 \mathbf{X}_p(t) = \frac{1}{N} \sum_{n=1}^{N} \mathbf{r}_n(t) \cos \left[ \frac{(n-0.5)p\pi}{N} \right]
\end{equation}
where $p=0, \ldots ,N-1$ and $\mathbf{r}_n$ denotes the center of
mass of monomer $n$. The static mode amplitudes, $\langle X_p^2 \rangle$ of the different models are shown in Figure 1(a). A dependence of  $\sin^{-2}(p\pi/2N)$ $\approx$ $p^{-2}$ would conform to the Rouse prediction which essentially considers the distribution of the normal coordinates to follow the Gaussian distribution. However, owing to the chain stiffness systematic deviation is
observed for all the models studied (see also \mcpcite{Balacu}).
All the Brownian systems, BD1, BDRA and BDCO show the same static $p$-behavior (not shown).
The homogeneous Brownian chain (BD1) shows a dependence of $p^{-2.2}$
only for the first few  modes and then changes to a steeper exponent reported earlier. The Brownian chain systems behave effectively as a freely rotating chain (FRC). The analytical result \mcpcite{Kreer2001} for  the normal mode amplitudes is given by 
\begin{equation}\label{FRCeqn}
 \langle \mathbf{X}_p^2 \rangle = \frac{R^2}{8N(N-1)} \left[ \left(\sin^2\frac{p\pi}{2N}\right)^{-1} - \left(\gamma^2+\sin^2\frac{p\pi}{2N}\right)^{-1} \right]
\end{equation}
with
\begin{equation}
\gamma^2=-\frac{(1+\langle \cos\theta \rangle)^2}{4\langle \cos\theta \rangle}.
\end{equation}
A fit of this equation with $\langle \cos\theta \rangle=-0.45$ agrees well with the simulation data (see Figure 1(a)). This angle is close to the actually observed angle of $\langle \cos\theta \rangle=-0.53$.

The neat PEO system shows a $p^{-2}$ dependence for small modes up to $p=5$. Beyond this the relationship smoothly curves to an approximate dependence of $p^{-4}$. Similar violations have been reported before for 1,4-polybutadiene\mcpcite{Krushev} and n-C$_{100}$H$_{202}$.\mcpcite{Paul} However, the $p$-dependence for the high frequency modes is much stronger in case of PEO. The self-similarity of the Gaussian coil over a broad length scale is no longer fulfilled. The cause can be attributed to the preference of sequences of conformational states that are more coiled than others.\mcpcite{BoroLiIStatic}

The polymer electrolyte system also exhibits a $p^{-2}$ dependence for only the first few modes and crossing over sharply beyond $p \approx  20$ to a $p^{-5}$ dependence. Specially, for high modes the presence of cations in the proximity of some monomers leads to a localised shrinking of the affected subparts of the chains. Precisely, the cations enforce a helical like arrangement of the
ether oxygen atoms around themselves and thereby further reducing the extent of spatial occupation of a contiguous group of bound monomers. This is consistent with the reduction in $R^2$ and also supported by the absence of such observations for the highest modes in the BD1 system where there is no torsional potential and non-bonded interactions.  

\begin{figure}
  \includegraphics[width=0.8\linewidth]{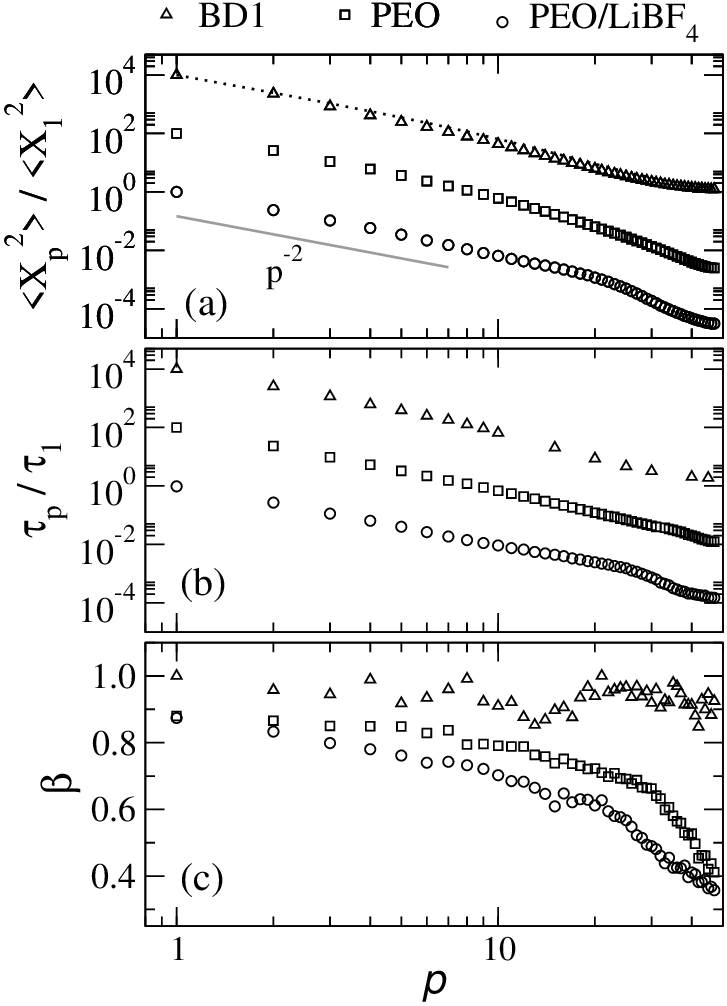}
  \caption{Analysis of Rouse modes. (Note that the data sets from the different models are offset for clarity in viewing) (a) Rouse mode amplitudes $\langle \mathbf{X}_p^2 \rangle$ scaled by the amplitude $\langle \mathbf{X}_1^2 \rangle$ of mode $p$=1 as a function of the mode index $p$. A fit using the Equation \eqref{FRCeqn} for a freely rotating chain with $\langle \cos\theta \rangle = -0.45$ is shown by the dotted line. (b) Scaled relaxation time of $p$th mode, $\tau_p$ by $\tau_1$ in dependence of mode index $p$. (c) Stretching parameter $\beta$ from the Kohlrausch-William-Watts (KWW) fitting, $\exp(-t/\tau_p)^\beta$, to the mode autocorrelation function $\langle \mathbf{X}_p(t)\mathbf{X}_p(0) \rangle$ as a function of $p$.
}\label{fig1}
\end{figure}

\subsubsection* {Dynamics}

Not surprisingly, the models BDRA and BDCO show biexponential decay of the autocorrelation functions of the modes because of the presence of two different kinds of particles in a chain. This effect becomes very pronounced with increasing mode index. A similar effect is expected for the PEO/LiBF$_4$ system, albeit diluted, due to frequent making and breaking of linkages with the ether oxygen atoms and intrachain cation motion.\mcpcite{BoroMDLiTFSI,MHprl} A non-exponential fit, like the Kohlrausch-William-Watts (KWW) fit, nevertheless, yields fair agreement for the microscopic models and BD1.

The Rouse theory predicts an exponential decay of the autocorelation function of the $p$-th mode:
\begin{equation}\label{corexp}
  \langle \mathbf{X}_{p}(t)\mathbf{X}_p(0) \rangle = \langle \mathbf{X}_p^2 \rangle \exp(-t/\tau_p).
\end{equation}
where $\tau_p$ is the time constant for the $p$-th mode. In general, deviation from the exponential behavior has been observed before.\mcpcite{Krushev,Smith_nongauss} We have extracted the parameters of the KWW form, $\exp\{-(t/\tau_p)^\beta\}$, where the  stretching parameter, $\beta$, characterises the non-exponentiality of the decay process.
Figure 1(b) shows the relationship between $\tau_p/\tau_1$ and the mode index $p$ for the different systems. The Rouse theory prediction of $\tau_p \propto p^2$ is observed only for the first few modes for the systems, reminiscent to the observation of the $p$-dependence of mode amplitudes.

\begin{figure}
  \vspace{-5pt}
  \includegraphics[width=0.9\linewidth]{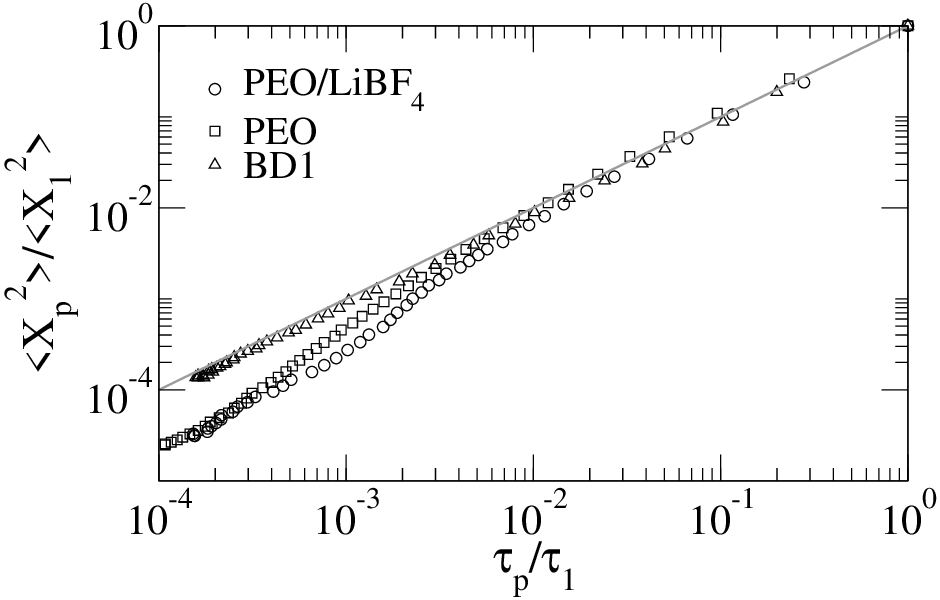}
  \caption{Scaled mode amplitude $\langle \mathbf{X}_p^2 \rangle$/$\langle \mathbf{X}_1^2 \rangle$ versus scaled relaxation time $\tau_p$/$\tau_1$ of the different modes obtained for PEO/LiBF$_4$ (circle), PEO (square) and BD1 models. The Rouse expectation $\langle \mathbf{X}_p^2 \rangle$/$\langle \mathbf{X}_1^2 \rangle$=$\tau_p$/$\tau_1$ is shown by the gray line.\label{fig2}
 }
\end{figure}

Figure 1(c) shows the relationship between $\beta$ and mode index $p$. For the Brownian chain, BD1, the $\beta$ values fluctuate around 0.9. Thus, the correlation functions for all modes are nearly exponential. The PEO and PEO/LiBF$_4$ systems show consistently decreasing values $\beta$ with increasing mode index. The non-exponentiality of the relaxation processes progressively
increases with decreasing length scales due to the existence of several structural realizations of a sequence of monomers in the presence and/or absence of a nearby cation. In general, the non-exponentiality of the relaxatory modes is stronger for the electrolyte than for the neat polymer due to the existence of cations in the vicinity of some monomers in the former.

The Rouse theory provides a linear relationship between $\langle \mathbf{X}_p^2 \rangle / \langle \mathbf{X}_1^2 \rangle$ and $\tau_p /\tau_1$. This is satisfied for the BD1 chain as shown in Figure 2 for all the modes. The agreement in case of PEO and PEO/LiBF$_4$ is only up to mode $p=12$ and $p=7$, respectively, and beyond which deviations are observed. The data suggest that these deviations may be related to the observed non-exponentiality in
$\langle \mathbf{X}_{p}(t)\mathbf{X}_p(0) \rangle$ which are beyond the scope of the Rouse theory.

\subsection* {Mean Square Displacement}
The mean square displacement (MSD), $g(t)$ of a polymer segment relative to the center of mass (c.o.m) of the polymer chain can be expressed in terms of its normal coordinates:
\begin{equation}\label{msd}
  g(t) = 4 \sum_{p=1}^{N-1} [ \langle \mathbf{X}_p^2 \rangle - \langle \mathbf{X}_p(t)\mathbf{X}_p(0) \rangle ]
\end{equation}

\begin{figure}
  \includegraphics[width=0.85\linewidth]{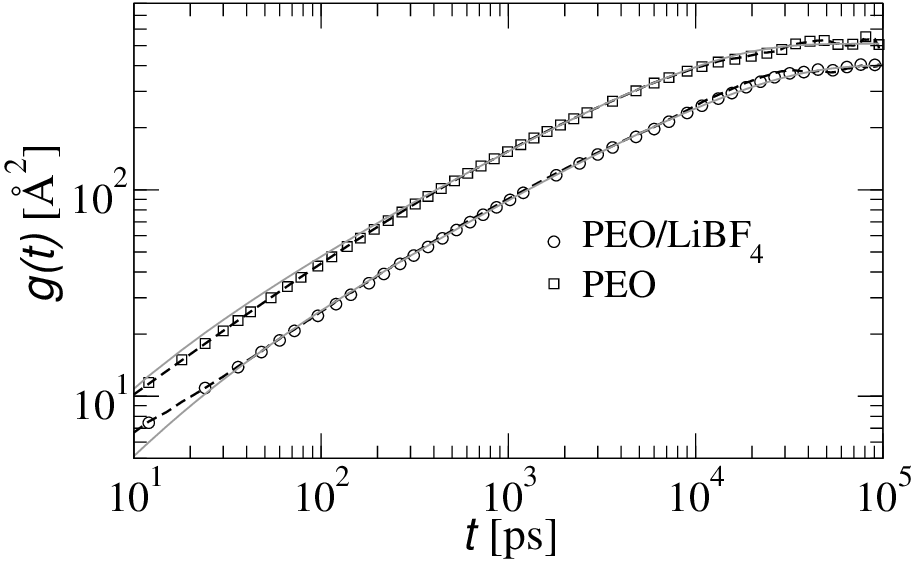}
  \caption{Mean square displacement (MSD) of ether oxygen against time for PEO/LiBF$_4$ (circle) and PEO (square). MSD calculated from Equation \eqref{msd} using the autocorrelation data extracted from simulations (dash). Rouse prediction of segmental dynamics (gray solid line). \label{fig3}
 }
\end{figure}

The only assumption entering Equation \eqref{msd} is the orthogonality of the modes.
Figure 3 shows the MSD of ether oxygen atoms of the PEO and PEO/LiBF$_4$ system. During the intermediate times, the segments experience sub-diffusional dynamics, $t^\alpha$, with $\alpha \approx 0.6$ arising from the topological constraints. Using Equation \eqref{msd} we have evaluated the MSD from the autocorrelation
function of the modes. The excellent agreement between the two, for both the systems, is indeed consistent with the orthogonality of the modes. We have also verified independently from the cross correlations (not shown) between the different modes that the modes are orthogonal to each other in agreement with similar observations before.\mcpcite{Kreer2001} The MSD from the ideal Rouse theory obtained by inserting Equation \eqref{corexp} into Equation \eqref{msd} and using $\tau_p=\tau_1/p^2$ is also compared in Figure 3. For the PEO system the theory and MD simulation results agree excellently beyond 100
ps. For shorter times, one sees only slight deviations. For the polymer electrolyte, beyond 20 ps the agreement with the Rouse prediction is very good. For times shorter than 20 ps the microscopic details, as captured by the non-exponentiality and the $p$-scaling of $\langle X_p^2 \rangle$, come into play and digressions from the ideal Rouse prediction become visible.

An ether oxygen atom is considered to be bound to a Li$^+$ if their separation is less than 3.0 \r{A}. We attempt to elucidate the difference in dynamics observed for bound and average (i.e. both free and bound) oxygen atoms (Figure 4) through the simplified Brownian chain models BDRA and BDCO. 

\begin{figure}
  \includegraphics[width=0.9\linewidth]{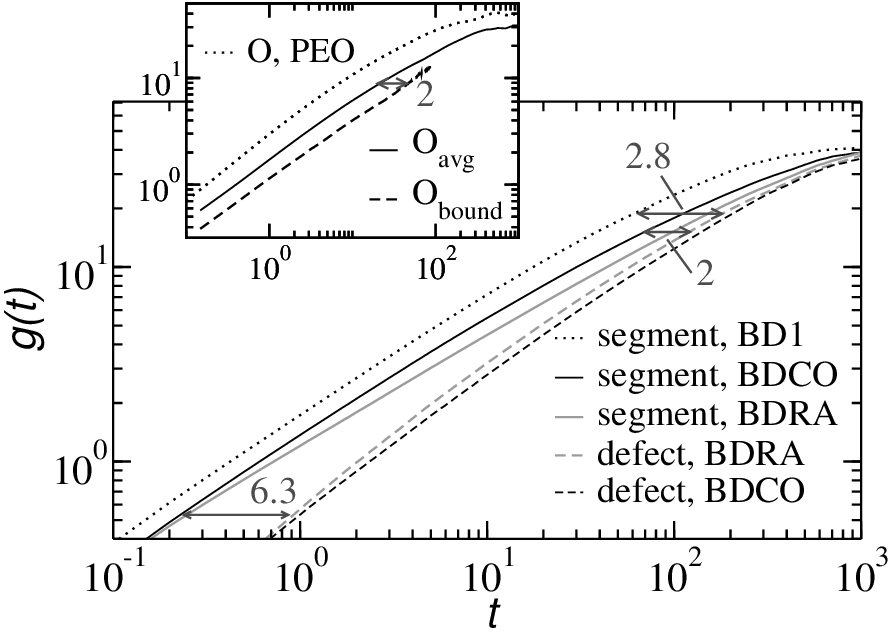}
  \caption{Mean square displacement (MSD) relative to chain center of mass: (i) all segments (solid) and (ii) defects (dash) computed for the toy models BDRA (gray) and BDCO (black), respectively. MSD of segments for the system BD1 are shown by dotted line. Also marked are the time scalings at short and intermediate times that is required for agreement between the relevant parts of $g(t)$ (see text for details). Inset: MSD of ether oxygen (EO) of neat PEO (dots); all EO atoms of PEO/LiBF$_4$ (solid) and EO atoms bound by Li$^+$ during time $t$ (dash); X and Y axes are scaled for the sake of comparison against the toy models.\label{fig4}
 }
\end{figure}

The MSD of the average segments and the defects with respect to the center of mass of the polymer chains are shown in Figure 4. For the BDRA as well as the
BDCO, at short times, when the particles are yet to feel the topological constraints, the difference in the dynamics can be expressed by the statistical factor:  g(all segments)/g(defect) =
$\zeta_2(0.75/\zeta_1+0.25/\zeta_2) \approx 6.3$. This is in agreement with the data; see Figure 4. 

At intermediate times, the average segment of BDRA is seen to lag behind that of the BDCO whereas the defect dynamics 
is slightly faster in case of BDRA compared to BDCO. 
Furthermore, the dynamics of the average segment and the defects merge with each other, corresponding to the value of $g(t) \approx R^2/3$, earlier for the BDRA than their counterparts for the BDCO. The reason is
that for BDRA the dynamics of the small-$p$ Rouse modes quickly gets averaged over the
fast and slow monomers. Slow and fast segments exchange their dynamics leading to retardation of the fast ones and acceleration of the slow beads until all segments share the same dynamics. This process occurs at a faster rate for BDRA as the possibility of each defect, on an average, to interact with non-defects is considerably higher than if the defects are spatially correlated. 

In contrast, for BDCO there exist extended regions of slow monomers which even at intermediate times can continue to move slower than the rest of the chain. Because of slower exchange of dynamics between the slow and fast segments, the process of averaging the segmental motion takes a longer time compared to the BDRA.

Additionally, we plot the MSD of the beads from the homogeneous chain (BD1). As expected, these show faster motion compared to the heterogeneous chains. At intermediate times, a time scaling of the MSD of the BD1 segments by a factor corresponding to the average friction coefficient of the heterogeneous chain $0.75\zeta_1 + 0.25\zeta_2 \approx 2.75$ (2.8 in Figure 4) is found to describe the defect dynamics of the BDRA chain fairly well. 

The dynamics of the EO atoms in the neat PEO is displayed in the inset of Figure 4. Also shown for the PEO/LiBF$_4$ system are the MSD of the average EO atoms and a subset of EO atoms that are bound to a Li$^+$ ion during time $t$. Note that the x and the y axes of all the PEO-based curves are scaled to match the units of the BD systems to facilitate a comparison. The oxygen dynamics in the neat PEO is significantly faster when compared to the average oxygen and the bound oxygen atoms of the polymer electrolyte. It can be observed that between $t=1$ and $t=100$ both the curves from the PEO/LiBF$_4$ system basically agree except for a time scaling with a factor of 2 (see inset of Figure 4). This time scaling signifies the relative immobilization of the bound oxygen atoms in the electrolyte.\mcpcite{MHprl} A similar factor of $\approx 2$ is observed in the BDCO toy system between the dynamics of the average segment and the defects in the time range of $t=50$ and $t=200$.
Thus, a  representation of the dynamics of the complexed monomers (or defects) necessitates an enhancement of the friction coefficients by one order of magnitude ($\zeta_2/\zeta_1$=8) to generate a right-shift of the dynamics of the average segments by a factor of $\approx 2$. This is also the reason behind choosing a value of $\zeta_2=8$ as the friction coefficient for the defects in our toy systems.

At intermediate times a time scaling by a factor of 3 relates the EO dynamics in the neat PEO to the PEO/LiBF$_4$ system. In contrast, a comparison of the average segmental dynamics of the BD1 and BDCO yields a factor of 1.4. This indicates that 
a rigorous comparison of EO dynamics in the neat PEO and the PEO/LiBF$_4$ is somewhat hampered owing to the differences in structure as well as the mean square end to end distance (of $\approx 20 \%$) between the two systems.
Nevertheless, the dynamics observed in BDCO is in qualitative agreement to the dynamics exhibited in the polymer electrolyte system.

\section* {Conclusions}
The polymer dynamics from all-atom simulations of PEO and PEO/LiBF$_4$ are expressed in terms of its normal coordinates or modes. The orthogonality of the modes holds and is supported by the accurate reproduction of the segmental dynamics of the monomers by taking into account only the autocorrelation functions of the modes and ignoring the presence of cross correlations. 
We have compared the findings to the predictions from the Rouse theory. The theoretical segmental dynamics considering only the exponential relaxation of the mode-autocorrelation function agrees with the monomer dynamics in case of PEO/LiBF$_4$ except for the
first 20 ps where the displacements are shorter than the statistical segment length. The agreement between theory and simulation holds for PEO only beyond 100 ps. This is surprising because one would have expected stronger deviations in case of polymer electrolytes than for neat PEO due to additional
generation of heterogeneities in the former by the complexed cations. For both PEO and the PEO/LiBF$_4$ the mode amplitudes show a dependence of $p^{-2}$ only for the first few modes before crossing over to a steeper dependence for high modes where local chemical details start to play, overriding the Gaussian
self-similarity seen over larger length scales. Furthermore, both the systems suffer from rising non-exponentiality with increasing mode numbers due to diversity in local environments and arrangement of conformational states. The violations of both $\langle \mathbf{X}_p^2 \rangle$ and $\tau_p$ from the $p^{-2}$ dependence is the cause of the steeper time exponents at intermediate times of the segmental dynamics compared to that predicted from Rouse theory. Finally, the effect of the presence of cations in the vicinity of segments augments the friction coefficients of the affected subparts of chain by an order of magnitude. Specifically, the dynamical effects of the
heterogeneities when they are confined along the chain in the form of contiguous units is more drastic than when they are distributed over the chain. Nevertheless, the Rouse model captures the average behavior of the segmental dynamics in PEO/LiBF$_4$ quite well. The dynamics of the bound segments can be obtained via simple time-scaling in the time regime which is relevant for the
understanding of the ion dynamics in polymer electrolytes.\mcpcite{MHprl} \\

\section* {Acknowledgement}
We gratefully acknowledge J. Baschnagel, O. Borodin and M. Vogel for helpful and valuable discussions. Furthermore, one of the authors (A.H.) would like to thank H. W. Spiess for sharing his insights into polymer dynamics and dynamic heterogeneities.



\end{document}